\documentstyle[mprocl]{article}

\def\apj#1{{\em Astrophys. J.} {\bf #1}}
\def\mn#1{{\em Mon. Not. R. astr. Soc.} {\bf #1}}

\def\prl#1{{\em Phys. Rev. Lett.} {\bf #1}}

\def\nat#1{{\em Nature,} {\bf #1}}
\def\apjs#1{{\em Astrophys. J. Suppl.} {\bf #1}}

\def\be{\begin{equation}}
\def\ee{\end{equation}}
\def\bea{\begin{eqnarray}}
\def\eea{\end{eqnarray}}
\def\etal{{\it et al.}\ }
\def\Mpc{$h^{-1}$~{\rm Mpc}}

\begin{document}

\title{LARGE SCALE STRUCTURE OF THE UNIVERSE\\
Introduction}

\author{J. EINASTO}

\address{Tartu Observatory, EE-2444 T\~oravere, Estonia}

\maketitle

\abstracts
{The changes of main paradigms on the structure and evolution of the
Universe are reviewed.  Two puzzles of the modern cosmology, the mean
density of matter and the regularity of the Universe on large scales,
as well as the possibility to solve these puzzles by the introduction
of more complicated form of inflation, are discussed.  }
  
\section{Changes of paradigms in cosmology}

Until mid-70's it was generally believed that galaxies form clusters
and groups, and the remaining field galaxies are more-or-less randomly
spaced in the Universe. In late-70's and early 80's this simple
picture was radically changed. It was found that field galaxies form
elongated chains or filaments, clusters also are often located along
chains; they form together filamentary superclusters of
galaxies~\cite{je78}. The space between such filaments is devoid of
any visible galaxies. This new picture was reviewed by Zeldovich,
Einasto and Shandarin \cite{zes82} and Oort \cite{oort83}.

The distribution of galaxies and clusters was confronted with
theoretical predictions by Zeldovich et al. It was found that this
distribution has some similarity with the distribution of particles
found in the adiabatic theory of structure formation developed by
Zeldovich and collaborators.  According to this scenario the structure
evolution is determined by the dominating dark matter population of
the Universe. If this population is due to massive neutrinos as it was
expected in early 80's, then perturbations on small wavelength are
damped, and the large-scale structural units, such as superclusters,
will form first. Matter flows from low-density regions which have
positive gravitational potential, to high-density regions forming
gravitation wells, and builds up pancake-like superclusters.  In
low-density environment the contraction of matter to form galaxies is
impossible, and the matter remains in some pre-galactic form.
Superclusters and voids form a continuous network of alternating high-
and low-density regions; the mean diameter of voids between rich
clusters of galaxies is about 100~\Mpc~\cite{zes82}.

Zeldovich et al. noticed also some problems with the neutrino
dominated Universe: in such picture only very rich superclusters form
and there are no systems of galaxies of intermediate richness; and, as
a result, voids should be completely empty. The observed structure is
more complicated: there exist intermediate sized systems of galaxies
that form rarified filaments between superclusters. This failure of
the neutrino-dominated Universe seems to be fatal, and it is overcomed
by a new candidate for the dark matter introduced by
Peebles~\cite{p82}. It is called cold since in contrast to hot
neutrinos particles of cold dark matter (CDM) have much lower
velocities~\cite{b84}. In CDM dominated Universe the formation of fine
structure is not damped and systems of galaxies of intermediate size
can form~\cite{m83}. All modern structure formation scenarios are
based on cold dark matter.

\section{Puzzles of modern cosmology}

The golden age of the theory of CDM Universe was 80's.  Numerical
simulations made within the standard CDM scenario with critical
density Universe were in much better agreement with observations than
simulations based on the HDM hypothesis~\cite{d85}. However, some weak
points in the standard scenario were found. It gives too low power on
large scales if normalised to small scales \cite{esm90}. The solution
of the problem was the introduction of models with a mixture of hot
and cold dark matter, or low $\Omega$ models with or without a
cosmological constant. These models can be characterised by the
parameter $\Gamma=\Omega h$ which determines the position of the
maximum and the power index of the spectrum on galactic scales.  The
standard model has $\Omega=1$ and $h=0.5$ which gives $\Gamma=0.5$; in
new models the preferred value is $\Gamma\approx 0.25$, hence for
$h\ge 0.6$ it follows that $\Omega \le 0.4$.  Direct dynamical density
estimates also support low density values. The case of a low-density
Universe with a non-zero cosmological constant was recently reviewed
by Ostriker and Steinhardt~\cite{os95}.

On the other hand, methods based on the study of the cosmic velocity
field yield higher values for the density parameter~\cite{d90}, and the
problem is still open for discussion. A number of talks in our
workshop are devoted to the discussion of the velocity field using new
data and methods of analysis.

Another cloud in the blue sky of the CDM-scenario has appeared
recently.  Superclusters and voids are formed by density waves of
wavelength which corresponds to the scale of the supercluster-void
network. According to the classical paradigm on the formation of the
large scale structure the distribution of density waves is Gaussian,
thus the distribution of high- and low-density regions should be
random.  It was a great surprise when Broadhurst \etal~\cite{beks90}
found that the distribution of high-density regions in a small area
around the northern and southern Galactic pole is fairly regular:
high- and low-density regions alternate with a rather constant step of
128~\Mpc. Bahcall and others \cite{b91,guz92} have confirmed that
these overdensities are part of extended supercluster-like structures.

This discovery rises the question: Has the Universe some regularity on
large scales, and if yes, what it means in terms of the structure
formation scenario?

Deepest available sources of information on the distribution of matter
on large scales are rich clusters of galaxies, catalogued by Abell and
collaborators~\cite{abell}, and the APM survey of galaxies and
clusters in the southern Galactic hemisphere~\cite{dmse97}.  Analyses
of these datasets are now available.

The 3-dimensional distribution of high-density regions as defined by
very rich superclusters of galaxies was found to be fairly regular
resembling honeycombs or 3-D chessboard~\cite{me94} with the same step
as found by Broadhurst.  This regularity can be described by the
correlation function of rich clusters of galaxies~\cite{e97b}, and by
the power spectrum of clusters~\cite{e97a}.  The cluster power
spectrum and the distribution of clusters in rich superclusters shall
be presented in my talk during this workshop.

The results of other independent analyses of Abell and APM clusters
also show the presence of a surprisingly sharp maximum in the power
spectrum~\cite{r97,t97}.  Three-dimensional reconstruction of the
power spectrum of 2-D distribution of galaxies of the APM survey
indicates again a rapid transition from the positive spectral index on
large wavelengths to negative index on galaxy scales
~\cite{p97,gaz97}.  The comparison of the power spectra with models
based on CDM scenario with the scale-free initial power spectrum has
shown that serious disagreement remains -- it is impossible to find a
set of cosmological parameters which yields a model in agreement with
new data on the power spectrum~\cite{r97,p97}.

Fluctuations of the temperature of the cosmic microwave background
radiation have been recently measured.  The peaked power spectrum of
matter determined from optical observations has been translated to the
angular CMB spectrum assuming a certain set of cosmological
parameters. Results of such comparisons~\cite{a97,Eis97} show that the
peaked power spectrum is in agreement with CMB data, but it cannot
identify a model of structure formation in an unique way.  Within the
framework of the classical scale-free initial power spectrum it is
extremely difficult to find a set of cosmological parameters that
satisfies all constraints.

\section{Is there light on the other end of the tunnel?}

We come to the conclusion that the present modernised CDM model of the
structure formation is in serious trouble. However, it seems that the
situation is not hopeless. All CDM models considered so far are based
on the assumption that inflation produces a scale-free initial power
spectrum, $P_0 \sim k$. This simple hypothesis is not the only
possibility. Already more than ten years ago more complicated variants
of the inflation scenario were suggested which predict a
non-scale-free initial power spectrum~\cite{ks85}. One of such
variants suggested by Starobinsky~\cite{s92} was recently compared
with CMB observations~\cite{a97,lps97}. The results are promising --
with a non-scale-free post-inflational spectrum it is possible to
satisfy simultaneously constraints posed by optical and CMB
observations.

Presently a series of new experiments is planned, both on the Earth
and in space.  We all look forward to see the results of these
experiments that certainly will give us much more accurate data on the
power spectrum in different regions and epochs. Whatever the answers
to our questions are, one is sure: the spectra on large scales give us
information on the structure of the Universe in the earliest epochs of
its evolution.

\section*{Acknowledgements}

I thank Enrique Gaztanaga, Alexei Starobinsky, Alex Szalay, and Helen
Tadros for stimulating discussions.

\end{document}